\documentclass{PoS}
\usepackage{amsmath,amssymb,bm,longtable,mathrsfs,slashed}

\title{\begin{picture}(0,0)(0,0)%
   \put(250,75){\makebox(0,0)[l]{\textnormal{\normalsize KEK-CP-370, OU-HET-987, RBRC-1302}}}%
\end{picture}Axial U(1) symmetry and Dirac spectra in high-temperature phase of $N_f = 2$ lattice QCD}

\ShortTitle{Axial U(1) symmetry and Dirac spectra in high-temperature phase of $N_f = 2$ lattice QCD}

\author{\speaker{Kei Suzuki}$\, ^a$\thanks{E-mail: \email{kei.suzuki@kek.jp}} , Sinya Aoki$\, ^b$, Yasumichi Aoki$\, ^{a,c}$\thanks{current address: KEK$^a$ and RIKEN Center for Computational Science, Kobe 650-0047, Japan.}, Guido Cossu$\, ^d$, Hidenori Fukaya$\, ^e$, Shoji Hashimoto$\, ^{a,f}$ \ (JLQCD Collaboration)\\
        \llap{$^a$}KEK Theory Center, High Energy Accelerator Research Organization (KEK), Tsukuba 305-0801, Japan\\
        \llap{$^b$}Center for Gravitational Physics, Yukawa Institute for Theoretical Physics, Kyoto 606-8502, Japan\\
        \llap{$^c$}RIKEN BNL Research Center, Brookhaven National Laboratory, Upton, NY 11973, USA\\
        \llap{$^d$}School of Physics and Astronomy, The University of Edinburgh, Edinburgh EH9 3JZ, United Kingdom\\
        \llap{$^e$}Department of Physics, Osaka University, Toyonaka 560-0043, Japan\\
        \llap{$^f$}School of High Energy Accelerator Science, The Graduate University for Advanced Studies (Sokendai), Tsukuba 305-0801, Japan\\
        }

\abstract{%
The axial $U(1)$ symmetry in the high-temperature phase is investigated with $N_f = 2$ lattice QCD simulations.
The gauge ensembles are generated with M\"obius domain-wall fermions, and the overlap/domain-wall reweighting is applied.
We find that the $U(1)_A$ susceptibility evaluated from the spectrum of overlap-Dirac eigenvalues is strongly suppressed in the chiral limit.
We also study its volume dependence.
}

\FullConference{The 36th Annual International Symposium on Lattice Field Theory - LATTICE2018\\
		22-28 July, 2018\\
		Michigan State University, East Lansing, Michigan, USA.}

\begin{document}
\section{Introduction}\label{sec-1}
The axial $U(1)_A$ symmetry plays an unique role in quantum chromodynamics (QCD).
In the low-temperature phase, it is violated by the chiral anomaly and is closely related to topological excitations of background gluon fields, such as the instantons.
As an order parameter of the $U(1)_A$ symmetry breaking, the $U(1)_A$ susceptibility, $\Delta_{\pi - \delta}$, may be defined by a difference between the correlators of isovector-pseudoscalar ($\pi^a \equiv i \bar{\psi} \tau^a \gamma_5 \psi$) and isovector-scalar ($\delta^a \equiv \bar{\psi} \tau^a \psi$) operators: 
\begin{equation}
\Delta_{\pi-\delta} \equiv \chi_\pi - \chi_\delta \equiv \int d^4x \langle \pi^a(x) \pi^a(0) - \delta^a (x) \delta^a(0) \rangle, \label{eq:Delta_def}
\end{equation}
where $a$ is an isospin index.
In this work we consider two flavor ($N_f=2$) QCD.

In the high temperature phase, the (spontaneously broken) chiral symmetry is restored, while the restoration (or violation) of the $U(1)_A$ symmetry remains a long standing problem.
%This has been studied using analytic approaches \cite{Cohen:1996ng,Aoki:2012yj,Kanazawa:2015xna} as well as lattice QCD simulations at $N_f=2$ \cite{Cossu:2013uua,Chiu:2013wwa,Brandt:2016daq,Tomiya:2016jwr} and $N_f=2+1$ \cite{Bazavov:2012qja,Buchoff:2013nra,Bhattacharya:2014ara,Dick:2015twa}.
%In Ref.~\cite{Cohen:1996ng}, Cohen claimed that the $U(1)_A$ symmetry of massless $N_f=2$ QCD is restored when the contributions from the zero modes of Dirac eigenvalues (i.e. nontrivial topological sectors) can be ignored.
%As a result, the mesonic correlators for $\pi$, $\sigma$, $\delta$, and $\eta$ channels become degenerate.
The JLQCD Collaboration observed a restoration of the $U(1)_A$ symmetry above the critical temperature $T_c$ in $N_f=2$ lattice QCD~\cite{Cossu:2013uua,Tomiya:2016jwr}.
Because the $U(1)_A$ susceptibility is sensitive to any tiny violation of chiral symmetry on the lattice, the lattice fermion formalism maintaining the chiral symmetry, such as the overlap (OV) or domain-wall (DW) fermions, was applied.
In Ref.~\cite{Cossu:2013uua}, the $U(1)_A$ symmetry was investigated using the Dirac spectrum on gauge configurations generated with the dynamical OV fermions in a fixed topological sector, $Q=0$.
In Ref.~\cite{Tomiya:2016jwr}, the gauge configurations are generated with dynamical M\"obius domain-wall (MDW) fermions \cite{Brower:2005qw,Brower:2012vk}.
Since the Ginsparg-Wilson (GW) relation \cite{Ginsparg:1981bj} for the MDW fermion is slightly violated especially for larger lattice spacings \cite{Cossu:2015kfa}, we applied the DW/OV reweighting technique \cite{Tomiya:2016jwr}.
In this case, observables measured on the gauge ensembles with dynamical MDW fermions is reweighted to that on OV fermion ensembles, for which the GW relation is precisely satisfied.

In these proceedings, we report the recent results of the $U(1)_A$ symmetry above $T_c$ in $N_f=2$ lattice QCD simulations with finer lattice spacing than Refs.~\cite{Cossu:2013uua,Tomiya:2016jwr}.
In particular, we will newly define the $U(1)_A$ susceptibility subtracted the ultraviolet divergence and compare the results with and without the ultraviolet contribution.
Note that a part of our results has already been reported in Refs.~\cite{Aoki:2017xux,Suzuki:2017ifu}.
%Our numerical setup is updated, compared to that in the previous paper \cite{Tomiya:2016jwr}.
%A finer lattice spacing, $1/a=2.64 \, \mathrm{GeV}$ ($a \sim 0.075 \, \mathrm{fm}$), is used, which improves the GW relation of the M\"obius domain-wall fermion.

%----------------------------------------------------------------------------
\section{Simulation setup}\label{sec-2}

\subsection{$U(1)_A$ susceptibility on the lattice}\label{subsec-2-1}
In the continuum theory, the $U(1)_A$ susceptibility (\ref{eq:Delta_def}) for fermion operators with a mass $m$ can be rewritten as
\begin{equation}
\Delta_{\pi-\delta} = \int_0^\infty d\lambda\,\rho(\lambda) \frac{2m^2}{(\lambda^2+m^2)^2}, \label{eq:Delta_cont}
\end{equation}
where the Dirac eigenvalue spectral density is defined by $\rho(\lambda)=(1/V)\langle\sum_{\lambda'}\delta(\lambda-\lambda')\rangle$ with the Dirac eigenvalues $\lambda$ and the four-dimensional volume $V=L^3\times L_t$.
On the lattice, the $U(1)_A$ susceptibility for OV fermion operators may be given by \cite{Cossu:2015kfa}
\begin{equation}
\Delta_{\pi-\delta}^{\mathrm{ov}} =  \frac{1}{V(1-m^2)^2} \left< \sum_i \frac{2m^2(1-\lambda_i^{(\mathrm{ov},m)2})^2}{\lambda_i^{(\mathrm{ov},m)4}} \right> , \label{eq:Delta_ov}
\end{equation}
where $\lambda_i^{(\mathrm{ov},m)}$ is the $i$-th eigenvalue of the (hermitian) massive overlap-Dirac operator, and the lattice spacing is set to $a=1$.
If the GW relation is not exact, we need additional terms in Eq.~(\ref{eq:Delta_ov}) \cite{Cossu:2015kfa}.
In the following, we discuss the subtraction of two contributions to the $U(1)_A$ susceptibility: the chiral zero modes and the ultraviolet divergence. 

Eq.~(\ref{eq:Delta_ov}) includes the effect of nontrivial topological sectors from chiral zero modes: $\lambda_i^{(\mathrm{ov},m)} \approx \pm m$, where ``$\approx$'' implies possible small violation of the GW relation in our simulations (If the GW relation is exact, then $\lambda_i^{(\mathrm{ov},m)} = \pm m$).
After subtracting such zero modes, we define a modified $U(1)_A$ susceptibility:
\begin{equation}
\bar{\Delta}_{\pi-\delta}^{\mathrm{ov}} \equiv \Delta_{\pi-\delta}^{\mathrm{ov}} - \frac{1}{V(1-m^2)^2} \left< \sum_{0-mode} \frac{2m^2(1-\lambda_i^{(\mathrm{ov},m)2})^2}{\lambda_i^{(\mathrm{ov},m)4}} \right> . \label{eq:Delta_bar}
\end{equation}
%Such a subtraction of zero modes can be justified as follows \cite{Aoki:2012yj}.
If the GW relation is exact, the second term of Eq.~(\ref{eq:Delta_bar}) can be written as $2N_0/Vm^2$, where $N_0$ is the number of chiral zero modes.
$\langle N_0^2 \rangle$ is expected to scale as $O(V)$, so that $N_0/V$ as $O(1/\sqrt{V})$.
Therefore, the contribution from the exact zero modes vanish in the thermodynamic limit: $N_0/V \to 0$ as $V \to \infty$.
By subtracting this contribution, the infinite volume limit would be approached faster.

Next we comment on the ultraviolet divergence.
Eq.~(\ref{eq:Delta_cont}) in the continuum theory contains an the (logarithmic) ultraviolet divergence, and Eq.~(\ref{eq:Delta_ov}) on the lattice includes a large contribution from the lattice cutoff $\Lambda$, which is proportional to $m^2 \ln \Lambda$.
Therefore, $\Delta_{\pi-\delta}(m)$ at a valence quark mass $m$ can be parametrized as
\begin{equation}
\Delta_{\pi-\delta} (m) = \frac{a}{m^2} + b + c m^2 + \mathcal{O} (m^4), \label{eq:Delta_parametrized}
\end{equation}
where the first term is the contribution from the zero modes.
The second term is the $U(1)_A$ violation which we are interested in.
The third term represents the contribution of $m^2 \ln \Lambda$.
Here, in order to eliminate $a$ and $c$, and to extract $b$, we use $\Delta_{\pi-\delta} (m)$ at three different valence quark masses ($m_2<m_1<m_3$):
\begin{equation}
b \simeq \Delta_{\pi-\delta}^\mathrm{finite} \equiv \frac{(m_1^2+m_2^2)(m_1^2+m_3^2)}{m_3^2-m_2^2} \left[\frac{m_1^2 \Delta(m_1) -m_2^2 \Delta(m_2)}{m_1^4 - m_2^4} -\frac{m_1^2 \Delta(m_1) - m_3^2 \Delta(m_3)}{m_1^4-m_3^4} \right]. \label{eq:Delta_subt}
\end{equation}
Note that we can use this formula also for $\bar{\Delta}_{\pi-\delta} (m)$ as defined in Eq.~(\ref{eq:Delta_bar}).
In that case, after the contributions of the zero modes in $\Delta_{\pi-\delta}^\mathrm{finite}$ are already subtracted, the contribution proportional to $a/m^2$ is absent.

%The overlap-Dirac eigenvalues $\lambda_i^{(\mathrm{ov},m)}$ measured on the M\"obius domain-wall fermion ensembles may include fictitious (zero and nearzero) modes induced by the partially quenched approximation \cite{Tomiya:2016jwr}.
%They come from a mismatch between the eigenvalues and the fermion determinant, which determines the Boltzmann factor for the particular gauge configuration.
%After the DW/OV reweighting, the mismatch is resolved by giving negligible small reweighting factor for the gauge configuration that suffer from the fictitious modes.

\subsection{Numerical setup}\label{subsec-2-2}
Our simulation parameters are summarized in Table \ref{Tab:param}.
We use the lattice with the spatial size $L=24,32,48$ and the temporal length $L_t =12$ which corresponds to $T=220 \, \mathrm{MeV}$ at the lattice spacing, $1/a=2.64 \, \mathrm{GeV}$ ($a \sim 0.075 \, \mathrm{fm}$).
For quark masses, we generate the data at $am=0.001-0.01$ ($2.64-26.4 \, \mathrm{MeV}$).

\begin{table}[t!]
  \small
  \centering
\caption{Numerical parameters in lattice simulations.
$L^3 \times L_t $, $L_s$, $\beta$, $a$, and $m$ are the lattice size, length of the fifth dimension in the M\"obius domain-wall fermion,
gauge coupling, lattice spacing, and quark mass, respectively.
}
\begin{tabular}{cccccc}
\hline\hline
$L^3 \times L_t $ & $L_s$ & $\beta$ & $a$ [fm] & $T$ [MeV] & $am$ \\
\hline
$24^3 \times 12$ & 16 & 4.30 & 0.075 & 220 & 0.001   \\
$24^3 \times 12$ & 16 & 4.30 & 0.075 & 220 & 0.0025  \\
$24^3 \times 12$ & 16 & 4.30 & 0.075 & 220 & 0.00375 \\
$24^3 \times 12$ & 16 & 4.30 & 0.075 & 220 & 0.005   \\
$24^3 \times 12$ & 16 & 4.30 & 0.075 & 220 & 0.01    \\
\hline
$32^3 \times 12$ & 16 & 4.30 & 0.075 & 220 & 0.001   \\
$32^3 \times 12$ & 16 & 4.30 & 0.075 & 220 & 0.0025  \\
$32^3 \times 12$ & 16 & 4.30 & 0.075 & 220 & 0.00375 \\
$32^3 \times 12$ & 16 & 4.30 & 0.075 & 220 & 0.005   \\
$32^3 \times 12$ & 16 & 4.30 & 0.075 & 220 & 0.01    \\
\hline
$48^3 \times 12$ & 16 & 4.30 & 0.075 & 220 & 0.001   \\
$48^3 \times 12$ & 16 & 4.30 & 0.075 & 220 & 0.0025  \\
$48^3 \times 12$ & 16 & 4.30 & 0.075 & 220 & 0.00375 \\
$48^3 \times 12$ & 16 & 4.30 & 0.075 & 220 & 0.005   \\
\hline\hline
\end{tabular}
\label{Tab:param}
\end{table}

We use the tree-level Symanzik improved gauge action.
For the fermion part, we apply the MDW fermions \cite{Brower:2005qw,Brower:2012vk} with a smeared link.
An observable $\mathcal{O}$ measured on the MDW fermion ensembles is transformed to that of the OV fermion by the DW/OV reweighting technique \cite{Tomiya:2016jwr}:
\begin{equation}
\langle \mathcal{O} \rangle_{\mathrm{ov}} = \frac{ \langle \mathcal{O} R\rangle_{\mathrm{DW}} }{ \langle R\rangle_{\mathrm{DW}}},
\end{equation}
where $\langle \cdots \rangle_{\mathrm{DW}}$ and $\langle \cdots \rangle_{\mathrm{ov}}$ are the ensemble average with the MDW and reweighted OV fermions, respectively.
$R$ is the reweighting factor which is stochastically estimated on the MDW fermion ensembles.
This procedure reduces the violation of the GW relation remaining with the MDW fermions.
In Section \ref{sec-3}, we compare $\lambda_i^{(\mathrm{ov},m)}$ and $\bar{\Delta}_{\pi-\delta}^{\mathrm{ov}}$ on the MDW ensembles with and without DW/OV reweighting.

\section{Preliminary results}\label{sec-3}

\begin{figure}[t!]
    \centering
    \begin{minipage}[t]{1.0\columnwidth}  
            \includegraphics[clip,width=1.0\columnwidth]{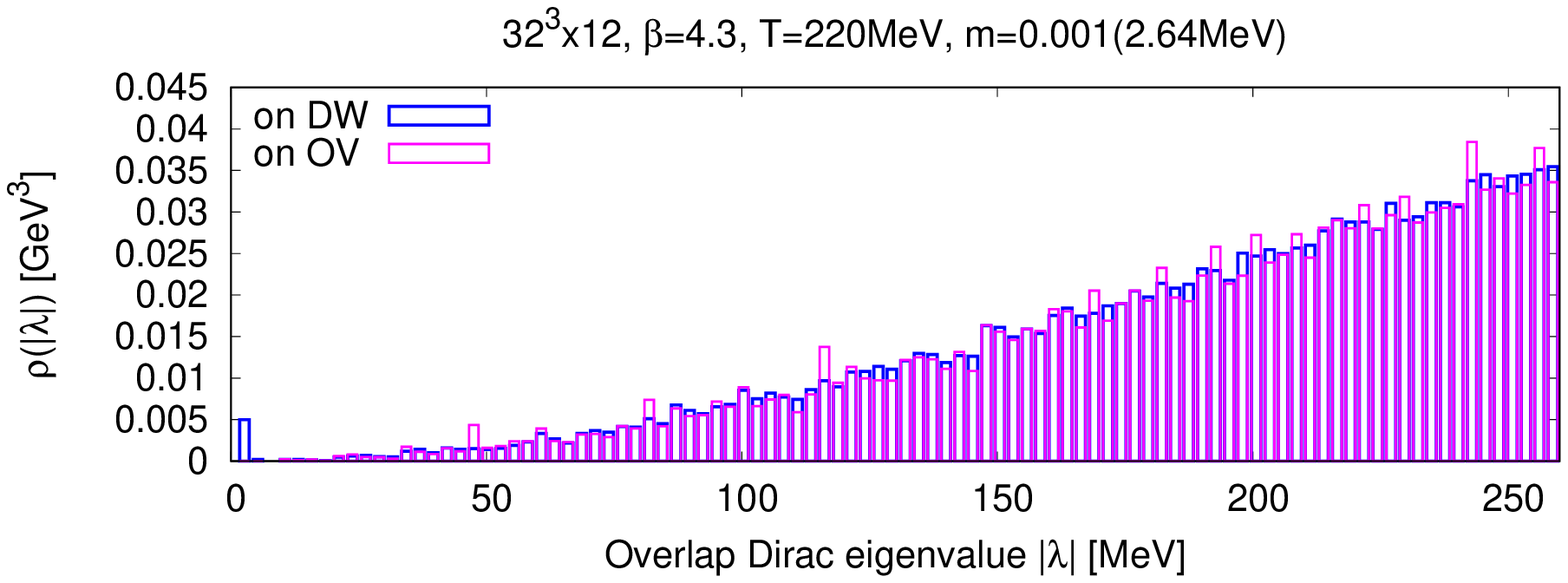}       
    \end{minipage}
    \begin{minipage}[t]{1.0\columnwidth}
            \includegraphics[clip,width=1.0\columnwidth]{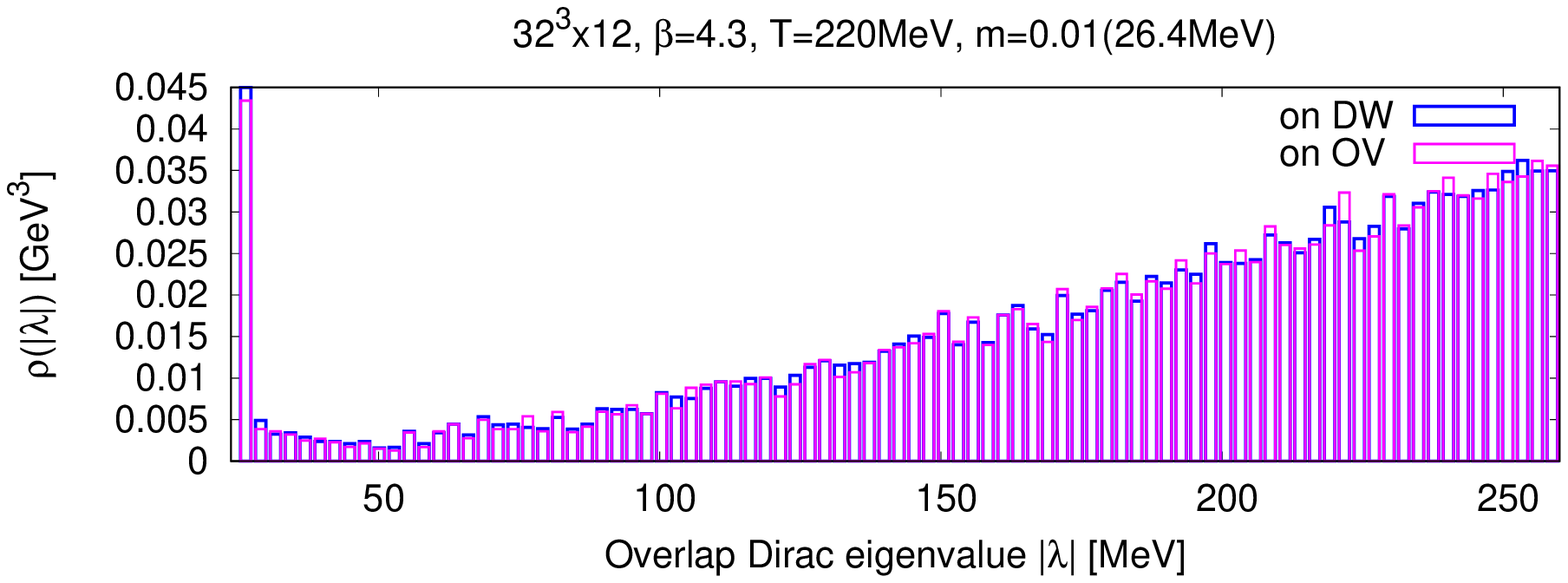}
    \end{minipage}
    \caption{Spectral density $\rho(|\lambda|)$ for overlap-Dirac eigenvalues $\lambda$ at $T=220 \, \mathrm{MeV}$.
Upper panel: $m=2.64 \, \mathrm{MeV}$.
Lower panel: $m=26.4 \, \mathrm{MeV}$.
Blue and magenta bins correspond to the spectra on the original M\"obius domain-wall (DW) and reweighted overlap (OV) fermion ensembles, respectively.
 }
    \label{fig-1}
\end{figure}

\subsection{Spectral density of overlap-Dirac eigenvalues}\label{subsec-3-1}
In Fig.~\ref{fig-1}, we show the spectral density $\rho(|\lambda|)$ of the overlap-Dirac eigenvalues $\lambda$ measured on the MDW ensembles at $T=220 \, \mathrm{MeV}$ with and without the DW/OV reweighting.
At the smallest quark mass $m=2.64 \, \mathrm{MeV}$ (the upper panel in Fig.~\ref{fig-1}), we find that the low lying eigenmodes are strongly suppressed.
We can clearly distinguish the zero modes from other higher modes.
Since Eq.~(\ref{eq:Delta_bar}) means that the finite value of $\bar{\Delta}_{\pi-\delta}^{\mathrm{ov}}$ comes from the non-zero modes, the suppression of the low modes is reflected in the small value of $\bar{\Delta}_{\pi-\delta}^{\mathrm{ov}}$.
Note that the zero modes in the spectrum of the DW (shown by blue bins) are likely due to artifacts caused by the mismatch between the valence and sea quark formulations (in other words, the partially quenched setup).
In the results after the reweighting (shown by magenta bins), such zero modes are removed.

At a larger quark mass (the lower panel in Fig.~\ref{fig-1}), the low-modes turned out to appear more frequently, and we cannot clearly separate the zero modes from other modes.
The increase of the low but nonzero modes leads to a large value of $\bar{\Delta}_{\pi-\delta}^{\mathrm{ov}}$.
In this case, the zero modes found on the DW ensemble survive even after the DW/OV reweighting, which indicates that such zero modes are not due to the lattice artifact.
% and can be related to the nontrivial topological sectors of QCD.

\subsection{$U(1)_A$ susceptibility}\label{subsec-3-2}
\begin{figure}[t!]
    \begin{minipage}[t]{0.5\columnwidth}
    \begin{center}
            \includegraphics[clip,width=0.94\columnwidth]{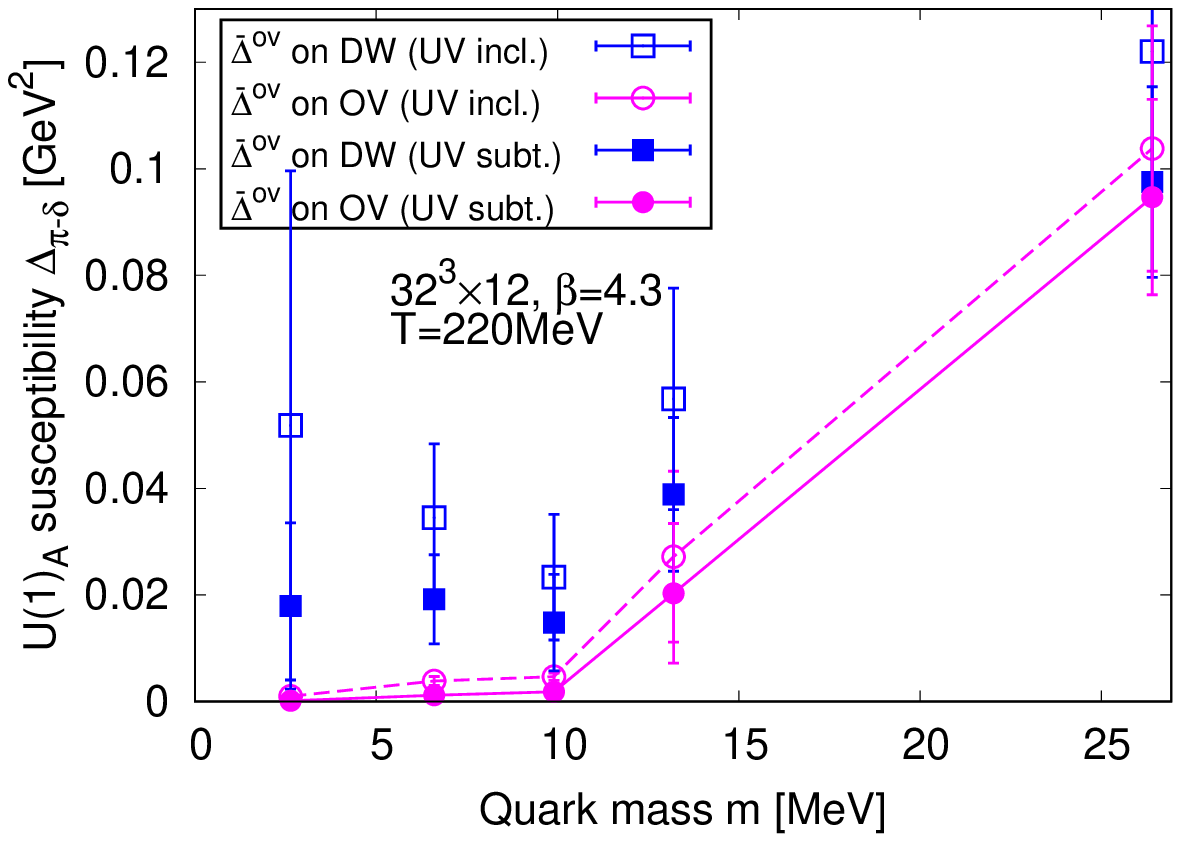}
    \end{center}
    \end{minipage}%
    \begin{minipage}[t]{0.5\columnwidth}
    \begin{center}
            \includegraphics[clip,width=1.0\columnwidth]{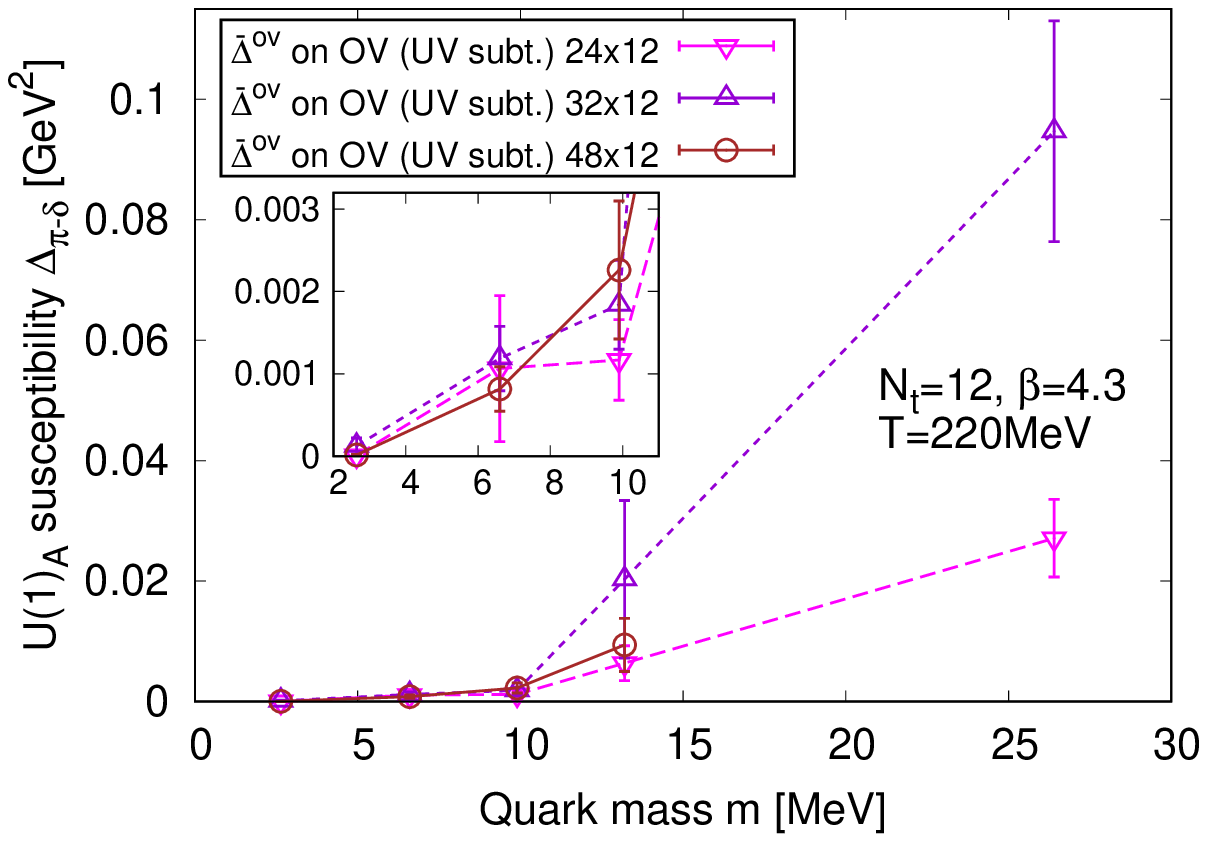}
        \end{center}
    \end{minipage}
    \caption{Quark mass dependence of $U(1)_A$ susceptibilities, $\bar{\Delta}_{\pi-\delta}^\mathrm{ov}$, from the eigenvalue density of the overlap-Dirac operators at $T=220 \, \mathrm{MeV}$.
Left: UV-included (open) and UV-subtracted (filled) results on the M\"obius domain-wall (squares) and reweighted overlap (circles) ensembles  at $L=32$.
Right: Volume dependence ($L=24,32,48$).}
    \label{fig-2}
\end{figure}

In Fig.~\ref{fig-2}, we show the quark mass dependence of the $U(1)_A$ susceptibility $\bar{\Delta}_{\pi-\delta}^{\mathrm{ov}}$ at $T=220 \, \mathrm{MeV}$.
The left panel shows the results at the spatial volume $L=32^3$.
Here, the circles (squares) correspond to the result on the OV (DW) ensemble.
Since $\bar{\Delta}_{\pi-\delta}^{\mathrm{ov}}$ on the DW includes fictitious modes by the partially quenched approximation, we expect that $\bar{\Delta}_{\pi-\delta}^{\mathrm{ov}}$ on the OV (circle) is closer to the continuum limit.
Also, the open (filled) symbols represent the results before (after) the UV subtraction (\ref{eq:Delta_subt}).
The former includes the ultraviolet contributions, so that it overestimates the $U(1)_A$ susceptibility by the amount that depends on the lattice cutoff.
%As a result, the filled magenta circles are the most reliable.

In the small quark mass region ($m \lesssim 10 \, \mathrm{MeV}$), $\bar{\Delta}_{\pi-\delta}^{\mathrm{ov}}$ on the OV nearly vanishes.
It strongly suggests that the $U(1)_A$ symmetry is restored in the chiral limit.
Near $m\sim 10 \, \mathrm{MeV}$, we find a sudden increase of $\bar{\Delta}_{\pi-\delta}^{\mathrm{ov}}$.
This may suggest the existence of a ``critical mass'' as discussed in Ref.~\cite{Aoki:2012yj}.
%In the large quark mass region, $\bar{\Delta}_{\pi-\delta}^{\mathrm{ov}}$ has a large value, which indicates that the $U(1)_A$ symmetry is clearly broken.

In the right panel of Fig.~\ref{fig-2}, we show the volume dependence of the $U(1)_A$ susceptibility.
For the small quark mass, there is no visible volume dependence between $L=24$ and $48$.
On the other hand, at large quark mass $m\sim 25 \, \mathrm{MeV}$, we found a clear volume dependence.
Investigation of the reason for this behavior is ongoing.

%In Fig.~\ref{fig-3}, we show the results at the almost same temperature, $T=217 \, \mathrm{MeV}$, shown by triangle points, but on a coarser lattice ($a\sim 0.11 \, \mathrm{fm}$) in Ref.~\cite{Tomiya:2016jwr}.
%Although the coarser lattice leads to larger violation of the GW relation for M\"obius DW fermion \cite{Cossu:2015kfa}, we found that $\bar{\Delta}_{\pi-\delta}^{\mathrm{ov}}$ is suppressed for the small quark mass region.
%Therefore, both the finer ($a\sim 0.075 \, \mathrm{fm}$) and coarser lattices lead to the similar suppression of  $\bar{\Delta}_{\pi-\delta}^{\mathrm{ov}}$.

%In the right panel of Fig.~\ref{fig-2}, we show the results at $T=330 \, \mathrm{MeV}$ which is much larger than $T_c$.
%At this temperature the $U(1)_A$ susceptibility is highly suppressed at all the quark mass parameters we studied.
%For instance, at $m \sim 27 \, \mathrm{MeV}$, $\bar{\Delta}_{\pi-\delta}^{\mathrm{ov}}$ after the reweighting is $\sim 10^{-4} \, \mathrm{GeV^2}$.
%This behavior comes from the suppression of Dirac low modes by the appearance of a gap in the Dirac spectra at higher temperature.

%----------------------------------------------------------------------------
\section{Conclusion and outlook}\label{sec-4}
In this work, we studied the $U(1)_A$ symmetry above the critical temperature from $N_f=2$ lattice QCD simulations.
The quark mass dependence of the $U(1)_A$ susceptibility at $T=220 \, \mathrm{MeV}$ suggests the restoration of $U(1)_A$ symmetry in the chiral limit, which is consistent with the theoretical prediction of Ref.~\cite{Aoki:2012yj}.
As other observables to characterize the restoration of $U(1)_A$, we may analyze hadronic correlators (for the spatial meson correlators from our gauge configurations, see Refs.~\cite{Rohrhofer:2017grg,Rohrhofer:2018pey}).

The $U(1)_A$ symmetry at lower temperature near the chiral transition and $N_f=2+1$ simulations needs to be studied.
The previous studies in Refs.~\cite{Bazavov:2012qja,Buchoff:2013nra,Bhattacharya:2014ara,Dick:2015twa} suggested the violation of $U(1)_A$ symmetry, and the comparison between these works and our results should be also clarify. 

%----------------------------------------------------------------------------
\section*{Acknowledgment}\label{sec-5}
Numerical simulations are performed on IBM System Blue Gene Solution at KEK under a support of its Large Scale Simulation Program (No. 16/17-14) and Oakforest-PACS at JCAHPC under a support of the HPCI System Research Projects (Project ID:hp170061).
This work is supported in part by the Japanese Grant-in-Aid for Scientific Research (No. JP26247043, JP18H01216 and JP18H04484), and by MEXT as ``Priority Issue on Post-K computer" (Elucidation of the Fundamental Laws and Evolution of the Universe) and by Joint Institute for Computational Fundamental Science (JICFuS).
%----------------------------------------------------------------------------
\bibliographystyle{JHEP}
\bibliography{lattice2018}

%\begin{thebibliography}{99}
%\end{thebibliography}

\end{document}